**Microbial iron reduction under oxic conditions: implications for subsurface biogeochemistry**

Giulia Ceriotti[1,3], Alice Bosco-Santos[1,3], Sergey M. Borisov[2], Jasmine S. Berg[1]

[1]Institute of Earth Surface Dynamics, University of Lausanne, Switzerland
[2]Institute of Analytical Chemistry and Food Chemistry, Graz University of Technology, Austria
[3]These authors contributed equally

**Abstract**

Iron (Fe) reduction is one of Earth's most ancient microbial metabolisms, but after atmosphere-ocean oxygenation, this anaerobic process was relegated to niche anoxic environments below the water and soil surface. However, new technologies to monitor redox processes at the microscale relevant to microbial cells have recently revealed that the oxygen ($O_2$) concentrations controlling the distribution of aerobic and anaerobic metabolisms are more heterogeneous than previously believed. To explore how $O_2$ levels regulate microbial Fe reduction, we cultivated a facultative Fe-reducing bacterium using a cutting-edge microfluidic reactor integrated with transparent planar $O_2$ sensors. Contrary to expectations, microbial growth induced Fe(III)-oxide (ferrihydrite) reduction under fully oxygenated conditions without forming $O_2$-depleted microsites. Batch incubations highlighted the importance of the process at a larger scale, fundamentally changing our understanding of Fe cycling from the conceptualization of metal and nutrient mobility in the subsurface to our interpretation of Fe mineralogy in the rock record.

## 1. Introduction

Iron (Fe) is one of the major elements of the Earth's crust, and microorganisms have been cycling it for at least 3.0 billion years[1] between its oxidized ferric Fe(III) and more soluble reduced ferrous Fe(II) forms. Today most of the Fe on the planet's surface is found within minerals like Fe(III) oxides in rocks[2], sediments[3], and soils[4], where they behave as nutrient and metal scavengers due to their high adsorbent capacity. Therefore, Fe-oxide dissolution controls adsorbed metal and nutrient mobility and bioavailability[4,5]. In other words, reductive Fe dissolution drives crucial environmental processes such as soil organic matter turnover and contaminant spreading in the subsurface.

Many abiotic and biotic reactions reduce Fe(III) oxides producing aqueous Fe(II)[6], which is used as a proxy to infer the occurrence of Fe(III) reduction in environmental studies[7]. Microbial Fe(III) reduction is carried out by dissimilatory metal-reducing bacteria that can be obligate[8] or facultative[9] anaerobes. Facultative Fe-reducers are assumed to preferentially respire $O_2$ according to the established thermodynamic electron acceptor cascade[10]. Consequently, like all anaerobic metabolisms, microbial Fe(III) reduction is thought to be confined to anoxic ($O_2$-depleted) environments.

Although dissolved Fe(II) rapidly reacts abiotically with $O_2$ (the redox potential $E°'$ of the pair Fe(II)/ Fe(III) = +0.36 V)[11], several studies have measured Fe(II), as free ions or complexed with organic matter, in oxic natural environments and experimental setups[6,12-16]. These observations point towards the occurrence of Fe(III) reduction processes in oxic environments. One potential explanation could be the formation of anoxic microsites, $O_2$-depleted zones generated by microbial $O_2$ consumption[14,17-19]. Due to their small size and ephemeral lifetime of hours to days, anoxic microsites are not detected by traditional $O_2$ measurements that capture only the bulk redox conditions of a system[18,19].

Microbial reduction of Fe(III)-oxide minerals in the presence of $O_2$ has been recently proposed for Actinobacteria, Proteobacteria, and Cyanobacteria[12,15,20-23]. Still, definitive proof beyond $O_2$ bulk analytical measurements to exclude the formation of anoxic microsites is lacking[4,17-19,24,25]. Characterizing microbial habitats at the microscale is thus essential to building an accurate understanding of microbial ecophysiology and the limits to anaerobic respiration [26-28].

Building on the most recent advancements in $O_2$ sensing techniques, we implemented a cutting-edge experimental approach to shed light on the influence of facultative Fe(III)-reducing bacteria on Fe cycling under oxic conditions across multiple observational scales. The facultative Fe-reducer *Shewanella oneidensis* MR-1 was grown in a microfluidic reactor with poorly crystalline Fe(III) oxides (ferrihydrite) under aerobic conditions. Oxygen concentrations were monitored at the microbial scale (micron) with fully transparent planar $O_2$ sensors. Larger-scale dynamics were explored in well-mixed batch experiments. Our results provide evidence for microbially induced ferrihydrite reduction under oxic and circumneutral pH conditions without forming anoxic microsites. This observation fundamentally changes our understanding of microbial Fe cycling in natural systems.

## 2. Experimental Procedures

### 2.1. Bacterial strain and growth medium

*Shewanella oneidensis* MR-1 (*S. oneidensis*) is among the most well-studied facultative Fe-reducing bacteria isolated from natural lake sediments[9,29]. *S. oneidensis* grown from frozen stocks overnight under oxic conditions in 5 mL of Luria-Bertani (LB, Sigma Aldrich) broth at 30°C in an orbital shaker (180 rpm).

To maintain circumneutral pH, the experimental medium was prepared with 20 mM PIPES buffer (piperazine-N,N′-bis(2-ethanesulfonic acid), Thermo Scientific) and enriched with LB broth (1:10 v/v) to avoid nutrient and Fe limitation while mimicking the complexity of carbon sources often encountered in natural systems (e.g., marshlands, peat bogs, eutrophic lakes, soils and even in and around marine snow)[30]. Ferrihydrite was synthesized in laboratory[31] and stored in closed vials protected from light. The medium was enriched with ferrihydrite, with a final concentration of 2 mM, similar to those found in soils[6]. Inoculation of microfluidic reactors and batch experiments was performed at a 1:50 (v/v) ratio of *S. oneidensis* overnight culture to Fe-LB medium.

### 2.2 Experimental setups and analytical methods

We designed a microfluidic reactor (Figure 1A) to study ferrihydrite reduction. Oxygen concentrations were continuously measured by integrated non-invasive $O_2$ planar sensors. This setup enables the monitoring of anoxic microsite (from micron to millimeter scale) formation which is otherwise undetectable with bulk measurements. Batch incubations (Figure 1B) complement microfluidic experiments to explore microbial Fe(III) reduction dynamics at a larger scale.

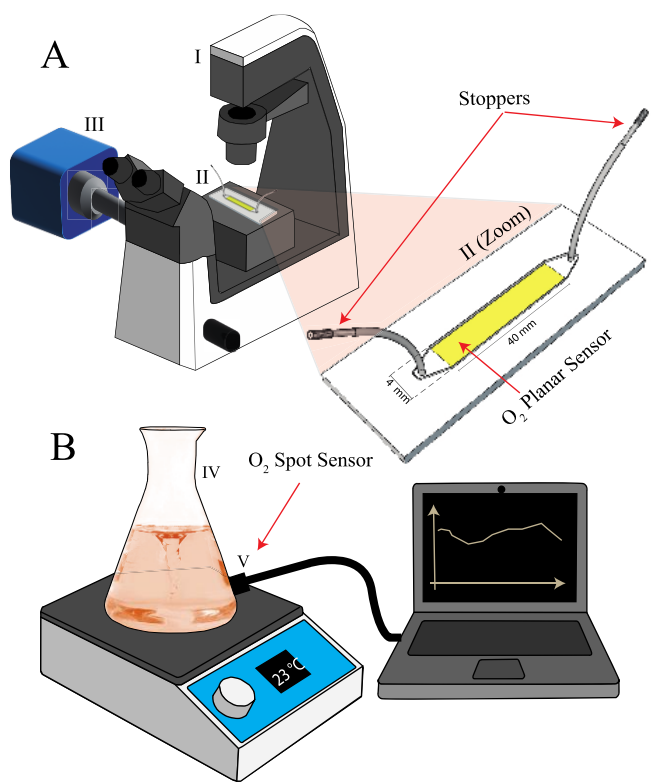

Figure 1. A- Illustration of the experimental setup used for the microfluidic incubation: an automated scope (I) hosting a microfluidic device shaped as a straight channel (II) on its stage and connected to a scientific camera (III) for capturing images. B- Sketch of the experimental setup used in well-mixed batch incubations: an Erlenmeyer flask (IV) integrated with a contactless $O_2$ spot sensor (V) taking measurements. The incubation solution was continuously stirred and maintained at a constant temperature.

**2.2.1 Microfluidic reactor setup (Figure 1A).** The microfluidic reactor shaped as a straight channel (see Figure 1A) of 40 mm (l) x 4.5 mm (w) x 100 μm (h) (total volume of ~18μL) was integrated with a fully transparent planar $O_2$ sensor[32] of 26 mm (l) x 4 mm (w)[24] (microfluidic fabrication and sensor calibration details are in SI, Section S1 and S2.1). Two Tygon tubes (Cole-Palmer, inner/outer diameter 0.02/0.04 inches) connected at the ends of the channel completed the microfluidic device.

After sterilization and degassing (see SI, Section S1), the microfluidic reactor was saturated with a freshly prepared medium inoculated with *S. oneidensis* (1:50 v/v) or the sterile medium as a negative control. Three replicates were performed for each condition. After saturation, the tube extremities were sealed by in-house produced stoppers (see SI, Section S1 for details) to prevent medium evaporation. The $O_2$ supply was ensured by diffusion from the atmosphere into the channel through the gas-permeable reactor walls[24].

To monitor $O_2$ concentration in real-time at the microscale in parallel with bacterial growth, an automated scope (inverted Nikon Eclipse Ti-E2) equipped with a 10X objective and a CMOS scientific camera (DS-Qi2 Nikon) took pictures (16-bit blank and white, 0.29 µm/pixel) of ~ 60% of the channel surface every hour for 72 h switching between different optical configurations (bright field, adjusted phase contrast, and two fluorescence configurations). Collected pictures were post-processed to map: a) ferrihydrite grains, b) microbial spatial organization and bulk growth, and c) $O_2$ concentrations. Technical details about the setup of the optical configurations and image processing are in the SI (Sections 3 and 4).

To measure Fe(II) concentrations in microfluidic reactors, we ran 5 identical microfluidic incubations in parallel. After incubating *S. oneidensis* for 168 h, the medium in the 5 parallel channels was extracted and mixed to collect sufficient liquid volumes (> 70 µL) for spectrophotometric measurements of Fe(II) using the Ferrozine assay[33] (see SI, Section S5 for details on method and calibration). The aforesaid procedure was replicated to obtain three independent measurements of Fe(II) after 168 h for the *S. oneidensis* incubations and the negative control (sterile medium). All microfluidic reactors were kept under strict dark conditions to rule out light-induced abiotic Fe(III) reduction.

**2.2.2 Well-mixed batch reactor setup (Figure 1B).** Batch incubations (*S. oneidensis* experiment, 3 replicates) were performed in 50 mL Erlenmeyer flasks filled with 25 mL of medium (without ferrihydrite) inoculated with 1:50 (v/v) of *S. oneidensis* overnight culture, as done in the microfluidic reactor. A magnetic stir bar ensured well-mixed conditions, $O_2$ exchange with the atmosphere, and temperature stability (23°C). After *S. oneidensis* attained stationary phase (72 h), ferrihydrite was added to the medium with a final concentration of 2 mM. Bulk $O_2$ concentrations in the experiment were monitored with $O_2$ sensor spots of 5 mm diameter (OXSP5 supplied by PyroScience, calibration procedure detailed in SI, Section S2.2) glued at the lower inner wall of the flask. Contactless measurements were logged in real-time at 6-minute intervals using the PyroScience Oxygen meter FireSting-$O_2$ connected to a laptop.

Every 24 h, an aliquot of 300 µL was sampled for Fe(II) concentration in solution and protein content as a proxy for microbial growth (Ferrozine and Coomassie assays, respectively; details in SI, Section S6). Measurements of pH were also taken to ensure circum-neutral conditions. To assess the possible contribution of abiotic processes in ferrihydrite reduction, two types of controls were run using the same ferrihydrite-enriched medium: *a)* negative control (three replicates) kept under sterile conditions (no inoculation), and *b)* dead controls (three replicates), where 37% formaldehyde (Sigma Aldrich) was added to a final concentration of 4% after inoculation to kill cells.

All the batch incubations ran continuously for at least 168 h in a sterile laminar-flow hood and sampled with sterile material. The experiments were performed completely in the dark to avoid photochemical reactions, which are known to contribute to Fe(III) reduction in acidic environments (pH < 6.5) directly, [34-38] or indirectly by boosting microbial metabolism[39].

## 3. Results

**3.1 Oxic conditions persist during aerobic microbial growth.** *S. oneidensis* was grown on LB with ferrihydrite particles in microfluidic reactors under oxic conditions. The spatial distribution of ferrihydrite, biomass, and $O_2$ concentrations was imaged simultaneously every hour (Figure 2A-F).

Although *S. oneidensis* densely colonized the microfluidic reactor (Figure 2A, C, E), $O_2$ concentration maps (Figure 2B, D, F) reveal that concentrations within the microfluidic reactor remained homogenous and close to air-saturation, between 8 – 8.4 mg/L (i.e., $O_2$ concentration in the air-saturated medium measured with TROX430 sensor, Pyroscience), throughout the experiment. We computed the percentage of microfluidic reactor space characterized by microoxic/anoxic conditions (i.e., with an $O_2$ concentration ≤ 0.32 mg/L)[28]. No microoxic/anoxic space was detected (see SI, Section S7 for procedures and data), confirming no anoxic microsites formed during *S. oneidensis* growth.

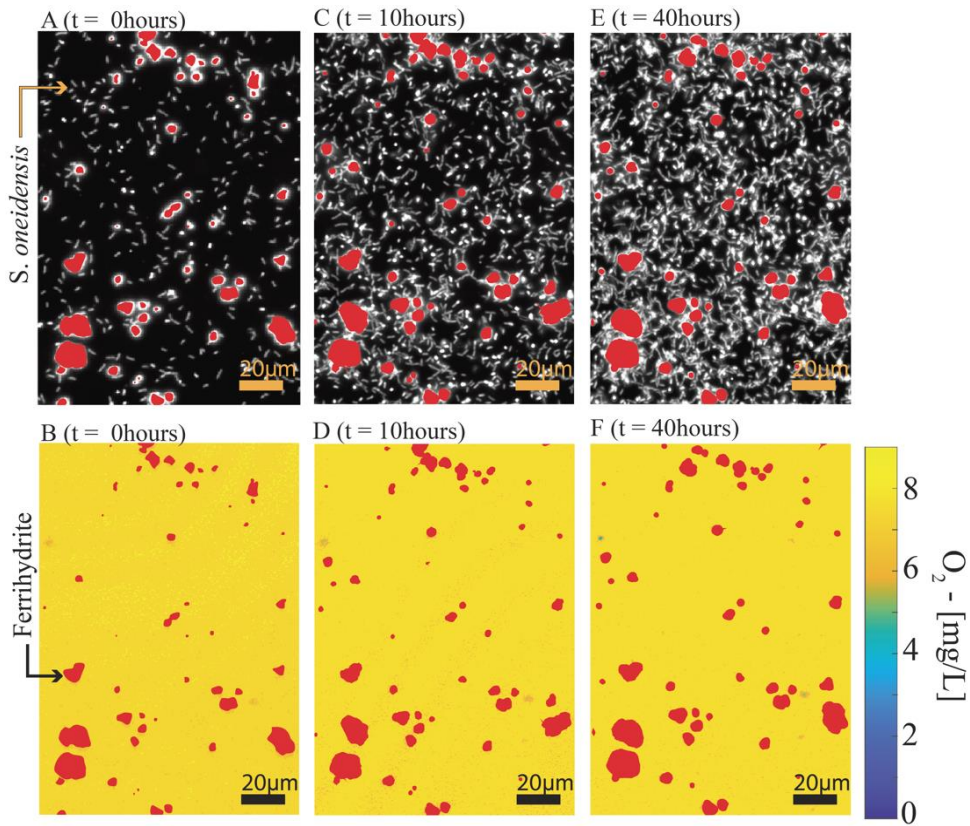

Figure 2 - Panels A, C and E: spatial distribution of ferrihydrite and *S. oneidensis* at t = 0, 10 and 40 h. Panels B, D and F: spatial distribution $O_2$ concentration and ferrihydrite at t = 0, 10 and 40 h, corresponding to panels A, C and E, respectively. The maps have a 0.29 µm/pixel resolution, or ~1/6th of a bacterial cell length.

To obtain a larger-scale perspective of the microbial activity in the reactor, bulk biomass growth ($BG_{Bulk}$, Figure 3A) and $O_2$ concentration ($O_{2Bulk}$, Figure 3B) were calculated by averaging pixel values in spatial maps (computational details in SI, Section S4). Exponential microbial growth within the first 24 h is consistent with the aerobic lifestyle of *S. oneidensis.* At the same time, bulk $O_2$ concentrations remained at 8.4 mg/L unaffected by microbial respiration and indistinguishable from the negative control. Microscale and bulk observations combined exclude the formation of anoxic microsites and indicate persistently oxic conditions. Given the time-invariant behavior of the system after 20 h, we assume that oxic conditions at the microscale are stable over longer time periods.

**3.2 Ferrihydrite reduction occurs under oxic conditions**. Despite the fully oxic conditions, dissolved Fe(II) was detected in the microfluidic reactors at the end of the 168 h incubation period. Experiments with live *S. oneidensis* exhibited significantly higher Fe(II) concentrations (133.7 ± 23.6 µM ; Figure 3C) compared to negative controls (12.8 ± 9.4 µM). These results suggest that S. *oneidensis* growth participates in ferrihydrite reduction under fully oxic conditions without the formation of anoxic microsites.

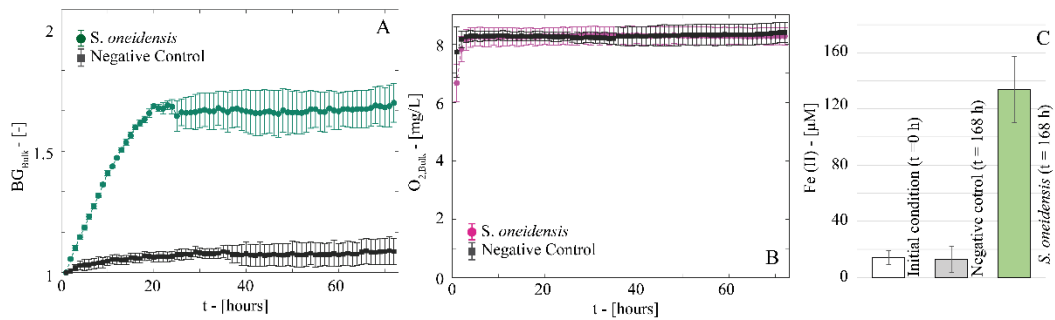

Figure 3 - Panel A: Temporal behavior of bulk microbial growth (BG$_{Bulk}$) measured as the average increment in light diffraction normalized to the initial value in the microfluidic reactors with *S. oneidensis* and the negative control. Panel B: Temporal behavior of bulk O$_2$ concentration computed as the average of spatial O$_2$ concentration maps. Panel C: Fe(II) concentration in the microfluidic reactors before and after 168 h of incubation.

**3.3 Oxic Fe(III) reduction is linked to living cells.** Fe(II) concentrations increased (Figure 4A) in larger-scale batch incubations of *S. oneidensis* upon the addition of ferrihydrite during the stationary phase of the bacterial growth (Figure 4B) while the recorded bulk O$_2$ concentration remained above 4 mg/L (data reported in SI, Section S8). These observations corroborate the observed trends of oxic Fe(III) reduction in microfluidic reactors.

Despite the presence of dissolved O$_2$, Fe(II) concentrations quadrupled within the first 24 h ($\Delta$Fe(II) = 260 µM, Figure 4A) after Fe addition (Figure 4B). By the 120$^{th}$ hour, average dissolved Fe(II) stabilized at 350 µM. Conversely, controls (negative and killed) showed no change in Fe(II) concentrations over time. These findings suggest minimal influence of abiotic processes and organic matter on Fe(III) reduction. To exclude the possibility that Fe(III) reduction is induced by nutrient scarcity, we successfully grew *S. oneidensis* in spent medium without ferrihydrite (SI, Section S9). Together, our results suggest that living cells, directly or indirectly, drive ferrihydrite reduction, independent of nutritional needs.

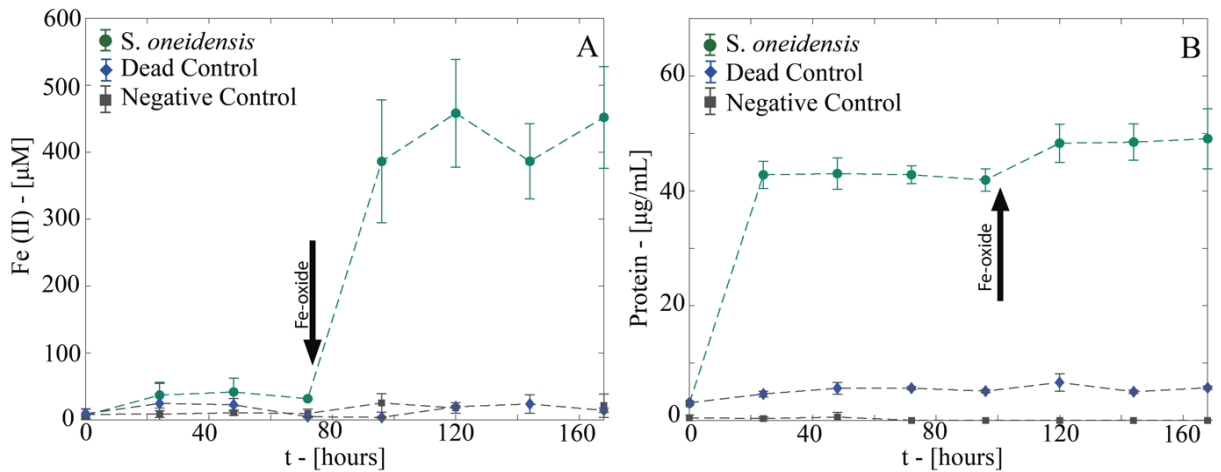

Figure 4 - Panel A: temporal behavior of Fe(II) concentrations in the batch *S. oneidensis* incubations, negative and dead controls. Sterile ferrihydrite was added to all of the batches at 73 h. Panel B: Protein concentrations as a proxy for microbial growth measured using the Coomassie assay in the same batches of panel A. All of the incubations were run in the dark.

## 4. Discussion

**4.1 A paradigmatic shift in biogeochemistry?** Traditionally, facultative Fe-reducing bacteria are predicted to use Fe(III) as a terminal electron acceptor only in the absence of $O_2$, which has a higher redox potential[40,41]. The terminal respiratory enzymes of *S. oneidensis,* in particular, are active under $O_2$ concentrations as low as 0.03 mg/L.[42-45]

Contrary to expectations, net Fe(II) production in our oxic *S. oneidensis* cultures demonstrates that Fe(III) reduction is viable under a wide range of $O_2$ concentrations. This challenges the traditional paradigm that Fe(III) reduction operates primarily under anoxic conditions assuming that bacteria preferentially utilize the electron acceptors yielding the most energy ($O_2$ > Fe(III)).

It was suggested as early as the 1990's that *Shewanella putrefaciens* sp. strain 200 could reduce Fe(III) under oxic conditions for unknown reasons[12]. However, the $O_2$-sensing technology employed at the time was limited to bulk $O_2$ concentrations, possibly overlooking the development of anoxic microsites inside bacterial aggregates. Here, we overcome this technological limitation with a new experimental microfluidic design allowing for real-time monitoring of $O_2$ concentrations at the microscale. We provide definitive proof of Fe(III) reduction under oxic conditions in the absence of anoxic microsites.

In live aerobic cultures, at least 11% of the original ferrihydrite was reduced, whereas in anaerobic cultures, Fe(III) was depleted by the end of the incubation (SI,

Section 10). In oxic batch reactors, the average Fe reduction rate observed after the addition of ferrihydrite ($15 \times 10^{-6}$ $M_{Fe}$ $h^{-1}$) falls within the range measured for anoxic incubations (from 6 to $42 \times 10^{-6}$ $M_{Fe}$ $h^{-1}$, see Figure. S4, SI). Together, this means that, although Fe mineral transformations are seemingly slower and less complete in the presence of $O_2$, they might not be irrelevant. These rates are slower than Fe(III) reduction by *S. putrefaciens*, $1.2 \times 10^{-5}$ M $h^{-1}$ and $24 \times 10^{-5}$ M $h^{-1}$ under oxic and anoxic conditions, respectively[12], and can be attributed to differences in Fe(III) bioavailability. Ferrihydrite used in our cultures to mimic minerals in natural systems is much less accessible than dissolved, complexed Fe(III) employed by Arnold, et al. [12]. Although Fe(III) oxide reduction under oxic conditions is clearly slower than under anoxic ones, the production of Fe(II) is significant in terms of bioavailability to other organisms and potential release of adsorbed nutrients and contaminants.

*Shewanella oneidensis* relies on Fe not only as an electron acceptor but also as a nutritional element integral to various cellular functions. Iron is involved in forming essential *S. oneidensis* cellular structures, particularly cytochromes[46]. Iron(III) reduction as a nutritional acquisition strategy has already been observed for cyanobacteria in the presence of $O_2$[13]. Although we could not definitively determine the primary purpose of Fe(III) reduction by *S. oneidensis*, we observed normal growth in spent medium without ferrihydrite (Figure S3, SI) indicating no nutrient (or Fe)-limitation. One possible explanation for Fe(III) reduction could be diversification of the respiratory chain as an adaptation to fluctuating environmental conditions. This means that either at the population or the cellular level, *Shewanella* retains the ability to respire both $O_2$ and Fe(III). This ecological strategy is already known for facultative denitrifiers that retain the enzymatic machinery for both aerobic and anaerobic respiration to quickly switch under fluctuating redox conditions[47].

Elucidating the reasons behind oxic Fe(III) reduction and the physiological mechanism behind it would require detailed genetic expression and enzymatic investigations that are beyond the scope of our current study. Nonetheless, a previous study of *S. putrefaciens* showed that stepwise addition of the aerobic respiration inhibitor CN⁻ to oxic cultures led to a gradual reduction in $O_2$ utilization without the ceasing of Fe(III) reduction[12]. The disparate rates of electron acceptor utilization point towards the involvement of a constitutive enzyme regulated independently of $O_2$ concentration[12].

**4.2 Persistence of a dissolved Fe(II) pool under oxic conditions.** During the batch incubations, we observed a relatively steady pool (average ~350 μM) of Fe(II) 48h after ferrihydrite addition (Figure 4). This observation contrasts with the long-standing belief that Fe(II) is extremely rapidly oxidized by free $O_2$ in pH-neutral conditions[6,48,49]. However, some previous reports indicate that dissolved Fe(II) may coexist with $O_2$[12,16,23,50].

In addition to ferrous Fe(II) and ferric Fe(III) ions, organic-Fe complexes (organic ligands such as humic and fulvic acids) are generally considered part of the dissolved Fe phases in natural systems[51,52]. In our experiments, dissolved Fe(II) is likely found as solvated Fe(II) ions or organic complexes within proteins from yeast extract (LB medium)[12]. Liganded organic-Fe(II) results in an "activated" phase that can serve as a reactive intermediate in biogeochemical processes. In practical terms, organic material may complex free Fe thereby affecting oxidation kinetics and result in the inhibition of hydrolysis and Fe(III) precipitation[53,54]. Complexation explains why, for example, noticeable Fe(II) concentrations persist in natural waters (e.g., soil porewaters [7,14] and rainwater[55]), and in controlled experiments[52]. In marine shelf sediments, such organic complexation could be responsible for diffusive Fe(II) fluxes (on the order of 23–31 μmol m$^{-2}$ day$^{-1}$) to the water column[53]. The same complexation mechanism coupled with continuous Fe(III) oxidation is therefore a possible explanation for the observed stable Fe(II) pool.

It is worth noting that a significant difference exists between the size of the Fe(II) pool persisting in the batch incubation, which is ~2.5 times larger than in the microfluidic reactor. This discrepancy is likely due to the stirring imposed in the batch that favors exchange of substrates and products and/or space limitation in the microfluidic reactor, leading to a lower number of active cells.

**4.3 Implications for subsurface biogeochemistry.** Due to its large specific surface area and poorly selective adsorption capacity[56,57], ferrihydrite often acts as a sink for a variety of anions[4,58] including hazardous metal(loid)s (e.g., arsenate, chromate, permanganate) and nutrients (e.g., phosphates). Reductive dissolution of ferrihydrite releases adsorbed anions and is therefore considered a key determinant of contaminant mobility. In fact, this process is reported to drive contaminant spreading and nutrient cycling in many anoxic environments such as flooded soils[59], deep groundwater systems[60,61], and anoxic microsites[14].

Our findings suggest that microbially induced Fe(III) reduction could play an unaccounted role in oxic environments. For example, microbial dissolution of ferrihydrite under oxic conditions could be the culprit of arsenate and chromate contamination observed in several oxic aquifers[62]. Similarly, ferrihydrite dissolution could occur in aerated, oxic soils, contributing to the surprisingly elevated content of mobile metals often observed in these systems[63].

It is worth noting that a large effort is nowadays invested in harnessing microbial Fe(III) reduction for remediation purposes. Fe(III) reducers are known to metabolize many recalcitrant contaminants while growing on ferrihydrite[64] with the added advantage of producing Fe(II) which is an effective reductant. Consequently, Fe reducers exhibit a potential for mediating the reductive immobilization of soluble toxic metals and the transformation of harmful compounds like chlorinated solvents[64,65]. Microbial Fe reduction under oxic conditions could be harnessed to develop new bio-remediation procedures also in shallow and well-drained subsurface systems.

Finally, our observations challenge the straightforward use of Fe valence state as a proxy for the redox conditions during mineral formation, opening a new perspective on our interpretations of the sedimentary Fe record. For instance, microbial Fe(III) reduction under oxic conditions might have contributed to the highly debated origin of the mixed-valence Fe(II)Fe(III)oxide, magnetite, in Banded Iron Formations (BIFs), without the need for diagenetic or metamorphic alterations.[66-71] The mineralogy of secondary Fe phases formed in our experiments (see SI, Section S10, Figure S6) will be the subject of future investigations.

## 5. Conclusions and future developments

In a cutting-edge setup complementing $O_2$-sensing microfluidic reactors with well-mixed batch cultures, we provide evidence for microbially induced ferrihydrite reduction in oxic aqueous environments. Using microscale spatial $O_2$ monitoring, we could exclude the contribution of anoxic microsites to Fe(III) reduction.

Microbial Fe(III) reduction in the presence of $O_2$ has relevant implications for mobilizing metals and nutrients in aquatic environments, such as shallow aquifers, soils, and lakes. Moreover, this microbially driven process presents an opportunity to reexamine the application of Fe valence states in minerals for reconstructing redox environment dynamics, especially within the context of the rock record.

Of course, to scale the relevance of Fe reduction in oxic natural subsurface systems, future investigations should characterize the environmental drivers of Fe(III) reduction rates (e.g., carbon source and Fe(III) concentration and type) and the impact of spatiotemporal soil/sediments heterogeneity. Our microfluidic approach offers a promising step in the direction of simulating more complex environmental scenarios, including soil-like porous structures, under fully controlled conditions[24]. From a microbiology perspective, the physiological mechanism behind microbial oxic Fe(III) reduction will require dedicated studies to elucidate.

## 6. Acknowledgments

The authors thank Prof. Rizlan Bernier-Latmani and Dr. Ashley Brown (Environmental Microbiology Laboratory, EPFL, Lausanne, CH) for providing the *Shewanella Oneidesis* MR-1 strain. The authors acknowledge Prof. Pietro de Anna (Environmental Fluid Mechanics Laboratory, UNIL, CH) for granting access to the microfluidic equipment and comments in the early stage of this investigation. Dr. Alice Bosco-Santos thanks the Agassiz Foundation for financial support (grant n. 26086987).

# Supplementary Information of: "Microbial iron reduction under oxic conditions: implications for subsurface biogeochemistry"

Giulia Ceriotti[1,3], Alice Bosco-Santos[1,3], Sergey M. Borisov[2], Jasmine S. Berg[1]

[1]Institute of Earth Surface Dynamics, University of Lausanne, Switzerland

[2]Institute of Analytical Chemistry and Food Chemistry, Graz University of Technology, Austria

[3]These authors contributed equally

## S1. Microfluidic reactor fabrication

We designed a straight channel shape as depicted in Figure S 1. The chosen design was printed onto a microfluidic master using classical soft lithography in a clean room with a final thickness of 100 μm. We used the microfluidic master to mold PDMS (Sylgard 184 Silicone Elastomer mixed with 10 w/w % of curing agent; supplier: Dow Corning, Midland, MI) and replicate the straight channel as many times as needed.

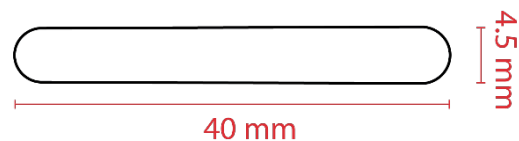

4.5 mm

40 mm

*Figure S 1. Design of the straight channel printed onto microfluidic master.*

The PDMS chip engraved with the straight channel shape was then assembled to a glass slide using plasma bonding. When needed, the PDMS chip was bonded to a glass slide equipped with an $O_2$ planar sensor, previously prepared as described in Section S2.

The two ends of the obtained microfluidic reactor were connected to 4-cm long Tygon tubing (Cole-Palmer, inner/outer diameter 0.02/0.04 inches).

After assembly, the microfluidic device is sterilized using UV light exposure for 30 minutes.

Before starting the incubation, the PDMS structure is degassed under vacuum for 20 minutes. Then, the straight channel was filled with the liquid medium using a sterile syringe, and each extremity of the tubing was sealed with a stopper. In a few minutes, air bubbles entrapped in the channel after medium injection were absorbed by the degassed PDMS structure.

The in-house produced stoppers were assembled by a 1cm-long Tygon tube half filled with NOA-81 glue (THORLABS) cured for 5 minutes under UV LED (365 nm, M365LP1, THORLABS). The cured glue created a fluid-impermeable barrier, acting as a stopper. The latter was connected to the Tygon tube extremities using a steel fitting (23G Fluid Adapter, Elveflow).

## S2. Sensor details and calibration

### 2.1 Oxygen planar sensor – microfluidic reactor

The transparent $O_2$ planar sensor integrated into the microfluidic reactor was obtained by screen-printing a homogeneous solution of two luminescent dyes in solid matrix polymer (polystyrene) onto a glass slide (75 mm x 25 mm) suitable for microscopy to obtain a homogenous layer with a thickness below 5 μm and shaped as a rectangle (26 mm x 4 mm).

The exact composition of the coating solution can be found in Ceriotti et al. (2022). The two dyes composing the sensor are excited by the same wavelength (450 nm) but show emission peaking at distinct wavelengths. The first dye (a phosphorescent Pt(II) porphyrin) emits a spectrum with a peak at 660 nm and its peak signal intensity ($I_{O_2}$) is quenched as a function of the $O_2$ concentration of the fluid in contact with the sensor. The second dye (fluorescent coumarin) is characterized by the emission spectrum that peaks at 500 nm. The peak signal intensity of coumarin dye ($I_{Ref}$) is insensitive to the oxygen concentration of the fluid in contact with the sensor.

The concentration of $O_2$ is given as a function of the ratio R = $I_{O2}$/ $I_{Ref}$ so that the sensor reading is purified from possible emission fluctuations caused by factors other than $O_2$ concentration variations (e.g., optical aberration, fluctuations in light intensity, etc.).

Optical configurations and video-microscopy technical details used for capturing $O_2$ planar sensor signal are reported in Section S3.

Sensor calibration was performed through a two-point procedure. Air-saturated calibration solution was prepared in a Schott bottle by shaking and exposing to air 10 mL of Fe-enriched medium three times. A similar amount of medium was amended with 100 mg of sodium sulfite ($Na_2SO_3$, Sigma Aldrich) to prepare the anoxic calibration solution ($O_2$ concentration = 0 mg/L).

The two calibration solutions were injected into a microfluidic reactor integrated with an $O_2$ planar sensor in sequence (starting with the air-saturated one). Images of the sensor luminescent signals were collected in 10 different locations along the channel length and post-processed (see Section S4 for procedures) to obtain the values of the ratio R corresponding to an $O_2$ concentration of air-saturated and 0 mg/L. Oxygen concentration in the air-saturated medium was fixed at 8.35 mg/L, measured with TROX430 sensor, Pyroscience, calibrated according to the guidelines.

Values of $O_2$ concentration and ratio R were interpolated with an exponential law following previous works using MATLAB Curve Fitting App (Ceriotti et al., 2022; Larsen et al., 2011).

$$O_2 \; concentration \left[\frac{mg}{L}\right] = 33.73e^{-2R} - 2.975$$

## 2.2   Oxygen spot sensors in batch reactors

We used oxygen spot sensors OXSP5 supplied by PyroScience. Sensors were connected to a laptop equipped with the software PyroWorkbench using the PyroScience Oxygen meter FireSting-$O_2$ (4 ports).

Spot sensor calibration was performed according to supplier guidelines, using the built-in two-point calibration procedure of PyroWorkbench and following guidelines.

## S3. Video-microscopy setup and optical configurations

Video-microscopy was performed with an inverted automated scope Eclipse Ti-E2, Nikon, controlled by NIS-Element software and equipped with i) a 10X objective; and ii) a CMOS DS-Qi2 (Nikon) with a sensor area of 36.0 mm x 23.9 mm and an actual pixel size of 7.3 μm and exposure time fixed at 200 ms. Images were taken at 10 locations every hour along the channel longitudinal direction, stitching together 4 pictures at each location to capture a more extensive area obtaining a final size of 9871 x 6365 pixels (corresponding to an area of ~ 5.2 mm$^2$) for each image.

The scope automatically switched between 4 different optical configurations:

- **Bright Field (BF).** This optical configuration used a diascopic scheme illuminating the device with the white LED light source at 0.5% of the maximum intensity. Opaque elements like ferrihydrite minerals were visible as dark objects on a white background.
- **Adjusted Phase Contrast (a-PC).** The same illumination source employed in BF was used at 64.3% of its maximum intensity combined with a Nikon Ph3 phase contrast plate compared to the ring (a Ph1) present in the selected objective. With this optical configuration, we detected all elements with an optical density different from the liquid medium, meaning that ferrihydrite minerals and microbial cells appeared as bright objects on a dark background.
- **Fluorescence at 500 nm ($F_{500}$) and 660 nm ($F_{660}$).** These optical configurations used an episcopic illumination scheme to capture the signals of the reference and the $O_2$-sensitive dyes, composing the $O_2$ planar sensor. A blue LED (440 ± 20 nm, Lumencor SPECTRA X Light Engine) illuminates the microfluidic reactor. This wavelength excited the luminescent dyes composing the $O_2$ planar sensor. The luminescent signal is filtered by Semrock bandpass emission filters (448 ± 20 nm for the reference and 650 ± 13 for the $O_2$ sensitive signals).

**S4. Image processing and computation of bulk parameters**

Image processing was performed in the Matlab environment with in-house produced codes.

**Ferrihydrite spatial organization.** The images captured with BF configuration were normalized to the highest pixel value recorded within each image so that each pixel attained a value between 0 and 1. The *pdf* (probability density function) of pixel value distribution and its $10^{th}$ percentile ($p_{10}$) were computed. All pixels with a value smaller than $p_{10}$ were identified as dark objects, i.e., ferrihydrite minerals. A binary matrix (*FE_OX*) with the same size as the original BF image for each acquisition time was generated to store ferrihydrite mineral spatial organization with a value equal to 0 for pixels associated with ferrihydrite minerals and 1 otherwise.

**Biomass spatial organization and bulk microbial growth.** Images collected with a-PC configuration were normalized to the maximum pixel value recorded in each image to obtain pixel values between 0 and 1. Pixels with higher values are associated with microbial and minerals, i.e., objects with an optical density more elevated than the liquid medium. These objects were identified by setting a threshold at 0.3 (chosen on preliminary assessment using images collected at t = 0 hours) and forcing all pixel values smaller than 0.3 to 0. To distinguish microorganisms from minerals, we multiplied the threshold a-PC image by the *FE_OX* matrix associated with the exact location and acquisition time. In the resulting map (*BIO*), only pixels associated with biomass presence showed values larger than 0.

Bulk microbial growth ($BG_{Bulk}$) was computed for each acquisition time ($t$) by processing the corresponding matrix *BIO* as follows

$$BG_{Bulk}(t) = \frac{\sum_{i=1}^{N} BIO(i,t)}{BG\_0 \; \sum_{i=1}^{N} FE\_OX(i,t)}$$

where $i$ is the image pixel counter, $N$ is the total number of pixels of the image, and $BG_{Bulk}$ ($t$=0) is equal to

$$BG\_0 = \frac{\sum_{i=1}^{N} BIO(i,t=0)}{\sum_{i=1}^{N} FE\_OX(i,t=0)}$$

**$O_2$ concentration maps and bulk $O_2$ concentration.** Spatial distributions of the ratio R were computed by dividing the images collected with $F_{660}$ configuration by those collected with $F_{500}$ one, pixel by pixel. $O_2$ concentration maps ($OX$) were obtained by applying the planar sensor calibration curve equation to each pixel value of R spatial maps.

Bulk $O_2$ concentration ($O_{2,Bulk}(t)$) was computed for each acquisition time ($t$) by processing the corresponding matrix $OX$ as follows

$$O_{2,Bulk}(t) = \frac{\sum_{i=1}^{N} OX(i,t)}{N}$$

## S5. Ferrozine assay – microfluidic reactor

A sample of > 70 μL was obtained by extracting and mixing the medium incubated in 5 replicates of the microfluidic reactor and treated in the following way:

- An aliquot of 80.6 μL of 0.1 M HCl was added to 70 μL of the sample in a 1.5 mL Eppendorf centrifuge conical tube to lyse bacterial cells and desorb Fe(II) from mineral surfaces during a 15 min reaction time;
- The solution was centrifuged for 15 minutes at 12000 rpm in a microcentrifuge to separate solids from the liquid phase;
- An aliquot of 140 μL of the supernatant is pipetted into a micro-well of a Greiner Microplate (96 wells, PS, F-bottom, clear) and mixed with 20 μL of Ferrozine reagent.
- After 10 minutes of reaction, the sample absorbance at 560 nm ($ABS_{560}$) was measured using a Spark® Multimode Microplate Reader.

The absorbance measurements are translated into Fe(II) concentrations by applying the following calibration curve

$$Fe(II) \ concentration \ [mM] = \begin{cases} 2.68 ABS_{560} - 0.122 & if \ ABS_{560} \leq 0.1 \\ 1.52 ABS_{560} - 0.014 & if \ ABS_{560} > 0.1 \end{cases}$$

The calibration curve was obtained by interpolating the absorbance measured for 8 calibration solutions (Fe(II) concentrations = 0, 0.005, 0.01, 0.015, 0.03, 0.1, 0.2, 0.5 mM) obtained by dissolving Fe-sulfate into the sterile medium, processed according to the same protocol applied to the samples. As $O_2$ can oxidize Fe(II), the determination of its concentrations might be underestimated (Porsch and Kappler, 2011; Posner, 1953).

## S6. Ferrozine and Coomassie assays – batch reactor

A sample of 300 μL was pipetted from the batch reactor with sterile material under the laminar-flow hood, transferred to a 1.5 mL Eppendorf centrifuge tube and mixed with 345 μL of 0.1 M HCl to lyse bacteria cells and desorb Fe(II) from mineral surfaces.

Protein concentration was measured as follows:

- After 15 minutes of reaction time with 0.1 M HCl, an aliquot of 150 μL was pipetted into a micro-well of a Greiner Microplate (96 wells, PS, F-bottom, clear) and mixed with 150 μL of Coomassie brilliant blue G-250.
- After 10 minutes of reaction, the sample's absorbance at 595 nm ($ABS_{595}$) was measured using a Spark® Multimode Microplate Reader.
- Protein concentration was computed through the following calibration curve

$$Protein \ concentration \ \left[\frac{\mu g}{mL}\right] = 68.7 \ ABS_{595} - 18.1$$

The protein concentration calibration curve was obtained following the Thermo Scientific protocol.

The concentration of Fe(II) was determined according to the following procedure:

- After transferring the aliquot used for the Coomassie assay, the solution was centrifuged 15 minutes at 12000 rpm in a microcentrifuge to separate solids from the liquid phase.
- An aliquot of 280 µL of the supernatant was pipetted into a micro-well of a Greiner Microplate (96 wells, PS, F-bottom, clear) and mixed with 20 µL of Ferrozine.
- After 10 minutes of reaction time, the sample absorbance at 560 nm ($ABS_{560}$) was measured using a Spark® Multimode Microplate Reader.

The concentration of Fe(II) was computed from $ABS_{560}$ by applying the calibration curve

$$Fe(II)\ concentration\ [mM] = \begin{cases} 2.0 ABS_{560} - 0.088 & if\ ABS_{560} \leq 0.1 \\ 1.33 ABS_{560} + 0.0098 & if\ ABS_{560} > 0.1 \end{cases}$$

The calibration curve was obtained by interpolating the absorbance measured for 10 calibration solutions (Fe(II) concentrations = 0, 0.005, 0.01, 0.015, 0.03, 0.05, 0.1, 0.3, 0.5, 1 mM) obtained by dissolving Fe-sulfate into the sterile medium, processed according to the same protocol applied to the samples.

## S7. Definition of anoxic micro-site percentage and its detection limit

To assess the formation of anoxic microsites, each $O_2$ concentration map ($OX$) was characterized in terms of the percentage of anoxic surface ($P_{AS}$ [%]). In this work, we assumed to be *anoxic* the space characterized by an $O_2$ concentration ≤ 0.32 mg/L, a threshold traditionally associated with the onset of microoxic conditions (Berg et al., 2022). To this end, we generated a binary matrix ($ANOX$) with the same size as the $O_2$ concentration map as follows

$$ANOX(i) = \begin{cases} 1\ if\ OX(i) \leq 0.32\ \dfrac{mg}{L} \\ 0\ if\ OX(i) > 0.32\ \dfrac{mg}{L} \end{cases} for\ i = 1, \dots, N$$

where $i$ is the pixel counter, and $N$ is the total number of pixels of the $O_2$ concentration map. The resulting matrix $ANOX$ identified all the pixels where the onset of anoxic conditions was observed (i.e., with $O_2$ concentration ≤ 0.32 mg/L). The percentage of anoxic volume results from

$$P_{AS}(t) = \frac{\sum_{i=1}^{N} ANOX(i,t)}{\sum_{i=1}^{N} OX(i,t)} 100$$

Small impurities entrapped in the microfluidic reactor, burnt pixels of the camera sensor, and imperfections of the $O_2$ planar sensor surface may locally interfere with the reading of the $O_2$ concentration and generate background noise in the computation of $P_{AS}$. To estimate the order of magnitude of the $P_{AS}$ background noise, we saturated a microfluidic reactor equipped with the $O_2$ planar sensor with a sterile air-saturated medium solution (20 mM PIPES, 2 mM ferrihydrite and 10X LB with $O_2$ concentration = 8.4 mg/L). To prepare the air-saturated solution, we used the same procedure reported in Section S2. Images of the sensor luminescent signals were collected in 10 different locations along the longitudinal direction of the channel and post-processed to obtain the corresponding $O_2$ concentration maps, as described in Section S4. We computed the corresponding $P_{AS}$ by applying the procedure reported above. Results are reported in Table S1, along with their mean value.

*Table S 1 PAS values computed for an air-saturated sterile medium in 10 different locations and their mean value.*

| Image | 1 | 2 | 3 | 4 | 5 | 6 | 7 | 8 | 9 | 10 | Mean Value |
|---|---|---|---|---|---|---|---|---|---|---|---|
| $P_{AS}$[%] | $1.1 \times 10^{-4}$ | $2.0 \times 10^{-4}$ | $7.1 \times 10^{-6}$ | $1.4 \times 10^{-5}$ | 0 | $7.6 \times 10^{-3}$ | $2.1 \times 10^{-3}$ | $1.8 \times 10^{-4}$ | 0 | 0 | $1.0 \times 10^{-3}$ |

Based on this assessment, we defined the value of $P_{AS}$ = 1.0 x 10$^{-3}$ as a technical detection limit for identifying anoxic micro-site formation. Any value of $P_{AS}$ falling below the detection limit was considered indistinguishable from the background noise and, therefore, was not related to the actual formation of anoxic conditions.

## S8. Bulk O₂ concentration in the batch incubations

In the *S. oneidensis* batch incubations, $O_2$ concentrations are, on average, lower than in the negative controls. The lowest $O_2$ concentrations were measured between 0 and 24 h during exponential growth. This means that vigorous stirring of the air-exposed culture does not fully compensate for the $O_2$ uptake by *S. oneidensis*. Nevertheless, the $O_2$ concentration always remains at levels associated with oxic conditions (Berg et al., 2022). The addition of ferrihydrite at t = 72 h had no impact on the $O_2$ concentration.

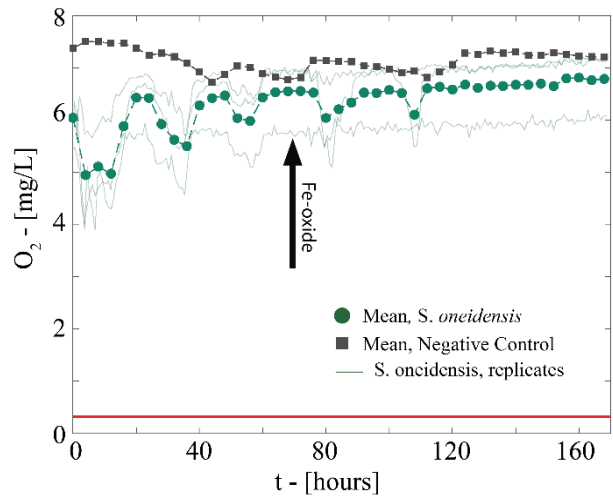

*Figure S 2 Measurements of O₂ bulk concentration for each replicate of well-mixed batch reactor with S. oneidensis and the corresponding mean value. Mean data are reported for the negative controls. The solid red line indicates the threshold for microoxic conditions and the black arrow identifies when ferrihydrite is added to the incubations.*

## S9.    Assessing Fe(II) limitation in the medium without ferrihydrite

In this experiment, we aimed to investigate the nutritional and, specifically, Fe limitations for the growth of *Shewanella oneidensis* under aerobic conditions in LB buffered medium. The experiments were conducted without the addition of Fe(III) oxides, in the dark, with continuous stirring over 145 h. Soluble Fe(II) was monitored at 24-h intervals (Figure S11-A). Simultaneously, the protein content, assessed by the Coomassie assay, was used as a proxy for bacterial growth (Figure S11-B).

After 72 h of incubation, the medium was filtered through a 0.22 μm filter. Subsequently, a fresh inoculum of *Shewanella oneidensis* was introduced into the sterile-filtered medium. The observation revealed that *Shewanella oneidensis* demonstrated renewed growth, indicating the absence of nutritional limitation.

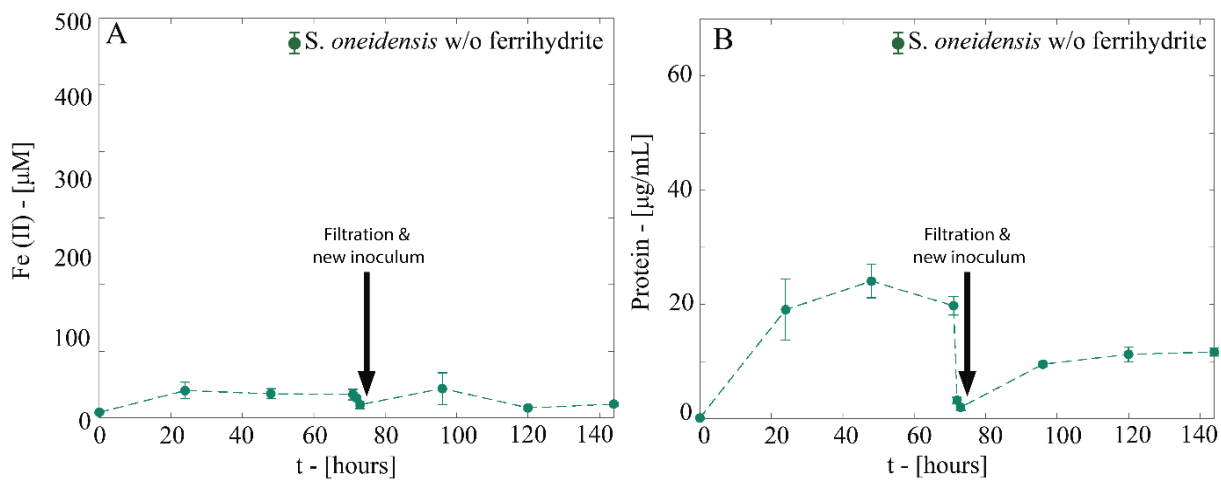

*Figure S 3 - Shewanella oneidensis* growth in LB buffered medium under aerobic conditions. Fe(II) concentrations (A) and protein content (B) were monitored over 145 h in the dark with continuous stirring. Inoculation at 0h, followed by sterile filtration and re-inoculation after 72 h, allowed renewed bacterial growth, indicating the absence of nutritional limitations in the culture medium.

## S10.  Iron reduction rate under oxic and anoxic conditions

To compare iron reduction rates under oxic and anoxic conditions, *S. oneidensis* batch incubations, along with negative controls, were run under $N_2$ (Figure S4) atmosphere.

Anoxic incubations were prepared in serum bottles with crimped butyl stoppers using the same incubation medium and inoculum of oxic experiments. Just after inoculation, batches were bubbled with $N_2$ gas filtered with a sterile 0.22 μm-filter to establish anoxic conditions. Samples of 300 μL were taken every 24 h with sterile syringes and flushed three times with $N_2$. Ferrozine and Coomassie assays were performed as reported in Section S6.

Fe(II) continuously increased, reaching a plateau at ~2 mM (Figure S3) when all of the Fe(III) is consumed. Microbial growth exhibited the expected logistic shape with a lag phase of 24 h and the stationary phase attained after 48 h.

Using the data reported here, we perform a preliminary assessment of the Fe(III) reduction rates under oxic and anoxic conditions.

In the first 96 h (see Figure 4), the average Fe(III) reduction rates ($6 \times 10^{-6}$ $M_{Fe}$ $h^{-1}$) under anoxic conditions are comparable to oxic conditions ($15 \times 10^{-6}$ $M_{Fe}$ $h^{-1}$) after the addition of ferrihydrite .

Under oxic conditions, it is not possible to estimate the reduction rate for a time larger than 24 h after the addition of ferrihydrite, given the stationary Fe(II) concentration. However, we suspect that Fe(III) reduction hasn't ceased, but it is compensated by rapid abiotic Fe(II) reoxidation, establishing a balanced Fe cycle. This cryptic cycling would also explain the significant visual change of the mineral aspect observed in the oxic incubation of *S. oneidensis*, not observed in the negative control (see Figure S6).

Under anoxic conditions, we observe an acceleration in Fe(III) reduction at 100 h ($42 \times 10^{-6}$ $M_{Fe}$ $h^{-1}$). We infer that, after a period of adaptation, *S. oneidensis* can perform Fe(III) reduction at considerably faster rates (at least 3 times) than those observed under oxic conditions.

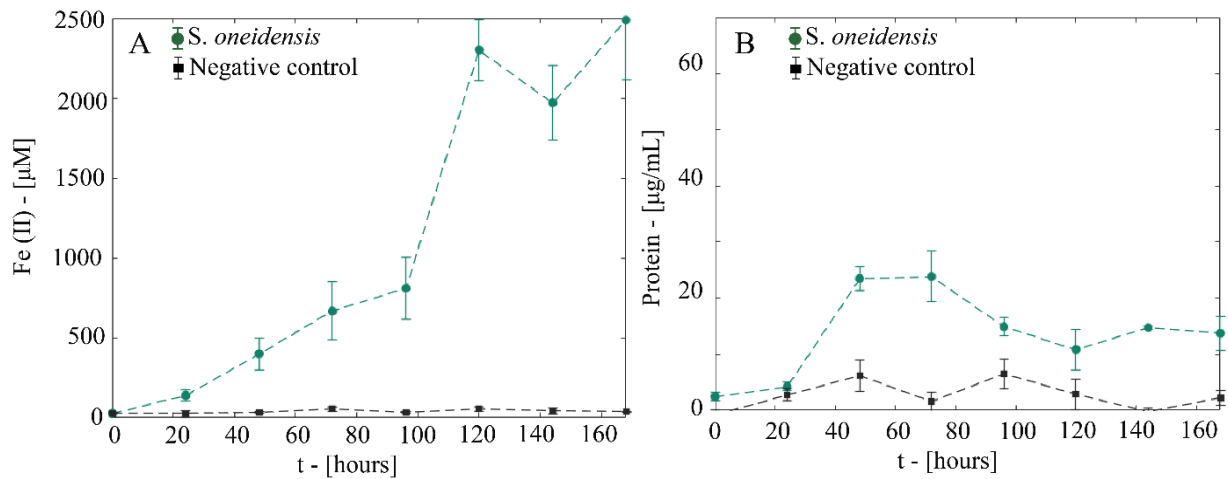

*Figure S 4 Panel A: Temporal behavior of Fe(II) concentration during S. oneidensis batch incubation and negative control under anoxic conditions. Panel B: Protein concentrations as a proxy for microbial growth measured as protein concentration using the Coomassie assay in the same batches of panel A.*

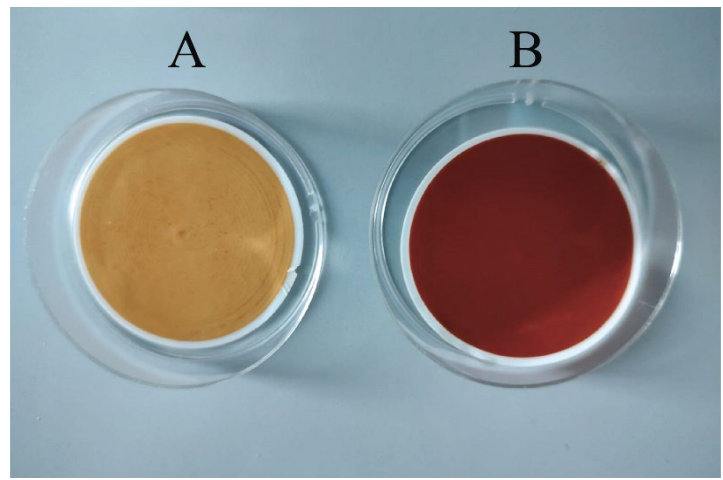

*Figure S 5 Visual assessment of mineral precipitates after 168 h of incubation with S. oneidensis (A) and in the negative control (B) under oxic conditions.*